\documentclass{article}   

\usepackage{graphicx}
  \usepackage{chicago}
  \usepackage[a4paper]{geometry}

\title{\bf Large Scale Structure Observations\footnote{Lectures given
    to the Fermi School on ``New Horizons for Observational
    Cosmology'', Varenna 1 - 6 July 2013, and to the Summer School
    ``Post-Planck Cosmology'', Les Houches 8 July - 2 August 2013}}
\author{ Will J. Percival\\
  \\
  \small{Institute of Cosmology and Gravitation, University of Portsmouth,}\\
  \small{Dennis Sciama Building, Portsmouth, P01 3FX, UK}}

\date{}

\begin{document}

\maketitle

\begin{abstract}
  Galaxy Surveys are enjoying a renaissance thanks to the advent of
  multi-object spectrographs on ground-based telescopes. The last 15
  years have seen the fruits of this experimental advance, including
  the 2-degree Field Galaxy Redshift Survey (2dFGRS;
  \shortciteNP{colless03}) and the Sloan Digital Sky Survey (SDSS;
  \shortciteNP{york00}). Most recently, the Baryon Oscillation
  Spectroscopic Survey (BOSS; \shortciteNP{dawson13}), part of the
  SDSS-III project \shortcite{eisenstein11}, has provided the largest
  volume of the low-redshift Universe ever surveyed with a galaxy
  density useful for high-precision cosmology. This set of lecture
  notes looks at some of the physical processes that underpin these
  measurements, the evolution of measurements themselves, and looks
  ahead to the next 15 years and the advent of surveys such as the
  enhanced Baryon Oscillation Spectroscopic Survey (eBOSS), the Dark
  Energy Spectroscopic Instrument (DESI) and the ESA Euclid satellite
  mission.
\end{abstract}

\tableofcontents

\section{Clustering statistics}

\subsection{The over-density field}  \label{sec:delta}

Any galaxy survey will observe a particular ``window'' of the
Universe, consisting of an angular mask of the area observed, and a
radial distribution of galaxies. In order to correct for a spatially
varying galaxy selection function, we translate the observed galaxy
density $\rho({\bf x})$ to a dimensionless over-density
\begin{equation}  \label{eq:delta}
  \delta({\bf x}) = \frac{\rho({\bf x})-\bar{\rho}({\bf x})}{\bar{\rho}({\bf x})}\,,
\end{equation}
where $\bar{\rho}({\bf x})$ is the expected mean density.

At early times, or on large-scales, $\delta({\bf x})$ has a
distribution that is close to that of Gaussian, adiabatic fluctuations
\shortcite{planck_ng}, and thus the statistical distribution is
completely described by the two-point functions of this field.

\subsection{The correlation function}  \label{sec:xi}

The correlation function is the expected 2-point function of this
statistic
\begin{equation}
  \xi(\mathbf{x_1},\mathbf{x_2})
    \equiv \langle
    \delta(\mathbf{x_1})\delta(\mathbf{x_2})
    \rangle\,.
\end{equation}
From statistical homogeneity and isotropy, we have that
\begin{eqnarray}
  \xi(\mathbf{x_1},\mathbf{x_2})
    & = & \xi(\mathbf{x_1}-\mathbf{x_2})\,, \nonumber\\
    & = & \xi(|\mathbf{x_1}-\mathbf{x_2}|)\,.
\end{eqnarray}
To help understand the correlation function, suppose that we have two
small regions, $\delta V_1$ and $\delta V_2$, separated by a distance
$r$. Then the expected number of pairs of galaxies with one galaxy in
$\delta V_1$ and the other in $\delta V_2$ is given by
\begin{equation}
  \langle n_{\rm pair} \rangle = \bar{n}^2\left[1+\xi(r)\right]
    \delta V_1\delta V_2\,,
  \label{eq:npair}
\end{equation}
where $\bar{n}$ is the mean number of galaxies per unit volume. We see
that $\xi(r)$ measures the excess clustering of galaxies at a
separation $r$. Imagine throwing down a large number of randomly
placed, equal length $r$, ``sticks'' within a survey. Then the
correlation function for that separation $r$ is the excessive fraction
of sticks with a galaxy close to both ends, compared with sticks that
have randomly chosen points within the survey window close to both
ends.

If $\xi(r)=0$, the galaxies are unclustered (randomly distributed) on
this scale -- the number of pairs is just the expected number of
galaxies in $\delta V_1$ times the expected number in $\delta V_2$.
Here, the same number of ``sticks'' have close-galaxies at both ends,
compared with randomly chosen points within the survey window.
$\xi(r)>0$ corresponds to strong clustering, and $\xi(r)<0$ to
anti-clustering. The estimation of $\xi(r)$ from a sample of galaxies
will be discussed in Section~\ref{sec:xi_est}.

\subsection{The power spectrum}  \label{sec:pk}

It is often convenient to measure clustering in Fourier space. In
cosmology the following Fourier transform convention is most commonly
used
\begin{eqnarray}
  \delta({\bf k}) & = & 
    \int\delta({\bf r})e^{i{\bf k.r}}d^3r\,,
    \label{eq:delr}\\  
  \delta({\bf r}) & = & 
    \int\delta({\bf k})e^{-i{\bf k.r}}
    \frac{d^3k}{(2\pi)^3}\,.
\end{eqnarray}
The power spectrum is defined as 
\begin{equation}
  P({\bf k_1},{\bf k_2})=\frac{1}{(2\pi)^3}
    \langle\delta({\bf k_1})\delta({\bf k_2})\rangle\,,
\end{equation}
with statistical homogeneity and isotropy giving
\begin{equation}
  P({\bf k_1},{\bf k_2})=\delta_D({\bf k_1}-{\bf k_2})P(k_1)\,,
\end{equation}
where $\delta_D$ is the Dirac delta function. The correlation function
and power spectrum form a Fourier pair
\begin{eqnarray}
  P(k) &= & 
    \int\xi(r)e^{i{\bf k.r}}d^3r\,,\\  
  \xi(r) & = & 
    \int P(k)e^{-i{\bf k.r}}
    \frac{d^3k}{(2\pi)^3}\,,
\end{eqnarray}
so they provide the same information. The choice of which to use is
therefore somewhat arbitrary (see \shortciteNP{hamilton1} for a
further discussion of this). In order to understand the power
spectrum, we need to think about ``throwing down'' Fourier waves,
rather than sticks, but the principle is the same. Methods to
calculate the power spectrum from a sample of galaxies are considered
in Section~\ref{sec:pk_est}.

\subsection{Higher order statistics} 

The extension of the 2-pt statistics, the power spectrum and
the correlation function, to higher orders is straightforward, with
Eq.~\ref{eq:npair} becoming
\begin{equation}
  \langle n_{\rm tuple} \rangle = \bar{n}^n\left[1+\xi^{(n)}\right]
    \delta V_1 \cdot\cdot\cdot \delta V_n\,.
\end{equation}
However, as we will discuss in Section~\ref{sec:pk}, at early times
and on large scales, we expect the over-density field to have Gaussian
statistics. This follows from the central limit theorem, which implies
that a density distribution is asymptotically Gaussian in the limit
where the density results from the average of many independent
processes. The over-density field has zero mean by definition so, in
this regime, is completely characterised by either the correlation
function or the power spectrum. Consequently, measuring either the
correlation function or the power spectrum provides a statistically
complete description of the field. Higher order statistics tell us
about the break-down of the linear regime, showing how the
gravitational build-up of structures occurs and allowing tests of
General Relativity.

\subsection{Anisotropic statistics}  \label{sec:aniso}

Although the true Universe is expected to be statistically homogeneous
and isotropic, the observed Universe is not so, because of a number of
observational effects discussed in
Section~\ref{sec:aniso_effects}. Statistically, these effect are
symmetric around the line-of-sight and, in the distant-observer limit
possess a reflectional symmetry along the line-of-sight looking
outwards or inwards. Thus, to first order, the anisotropies in the
over-density field can be written as a function of $\mu^2$, where $\mu$
is the cosine of the angle to the line-of-sight. Consequently, we
often write the correlation function $\xi(r,\mu)$ and the power
spectrum $P(k,\mu)$.

It is common to expand both the correlation function and power
spectrum in Legendre polynomials $L_\ell(\mu)$ to give multipole
moments,
\begin{equation}  \label{eq:Pl_def}
  P_\ell(k)\equiv\frac{2\ell+1}{2}\int^{+1}_{-1}d\mu\,P(k,\mu)\,L_\ell(\mu)\,.
\end{equation}
The first three even Legendre polynomials are $L_0(\mu)=1$,
$L_2(\mu)=(3\mu^2-1)/2$ and $L_4(\mu)=(35\mu^4-30\mu^2+3)/8$\,. 

\section{The comoving matter power spectrum}  \label{sec:pk_phys}

The present-day matter power spectrum is the evolved result of the
primordial power spectrum produced during inflation, a period of rapid
acceleration in the early Universe. Inflation explains why apparently
causally disconnected regions have similar properties, and why the
energy-density is close to the critical value, as well as explaining
the existence of present-day structure. While the general paradigm of
inflation is widely accepted, the details are largely unconstrained
(see \shortciteNP{byrnes10}). Although we know that the post-inflation
distribution of fluctuations has a statistical distribution close to a
Gaussian \shortcite{planck_ng}, determining the amount of
non-Gaussianity provides a key way of distinguishing between models.

The post-inflation matter power spectrum is commonly parameterised by
a power law, $P(k)\propto k^n$, where the power-law index $n=0.96$
\shortcite{planck_param}. This power spectrum is subsequently altered
by physical process within the evolving Universe composed of
radiation, baryons, neutrinos, dark matter and dark energy. In
particular, the relative densities of these components changes with
scale factor $a$ according to the Friedman equation, which for a
$\Lambda$CDM Universe can be written in the form
\begin{equation}  \label{eq:Frie}
  E^2(a)\equiv\frac{H^2(a)}{H_0^2}
    =\Omega_Ra^{-4}+\Omega_Ma^{-3}+\Omega_ka^{-2}+\Omega_\Lambda\,,
\end{equation}
where $E(a)$ is the Hubble Parameter $H(a)$ normalised to its present
day value, $\Omega_R$, $\Omega_M$, and $\Omega_\Lambda$ are the
present-day radiation, matter and $\Lambda$ densities in units of the
critical density, and $\Omega_k$ is the spatial curvature density
($=1-\Omega_{\rm tot}$). $a$ is the cosmological scale factor,
$a=1/(1+z)$, such that this equation and the parameters within can be
written as functions of either $a$ or $z$. The physics, including the
important processes that are described in the following subsections,
is usually encoded in a transfer function $T(k)$, which details the
change in the power spectrum from the inflationary form through to the
power spectrum in the matter dominated regime (the regime where the
dominant term in Eq.~\ref{eq:Frie} is $\Omega_Ma^{-3}$).

\subsection{The matter-radiation equality scale}

The growth of dark matter fluctuations is intimately linked to the
Jeans scale. Perturbations smaller than the Jeans scale do not
collapse due to pressure support -- for collision-less dark matter
this means being supported by internal random
velocities. Perturbations larger than the Jeans scale grow through
gravity at the same rate, independent of scale. If we approximate the
Universe as containing just dark matter and radiation, the Jeans scale
grows to the size of the horizon at matter-radiation equality, and
then reduces to zero when the matter dominates. We therefore see that
the horizon scale at matter-radiation equality will be imprinted in
the distribution of fluctuations -- this scale marks a turn-over in
the growth rate of fluctuations. What this means in practice is that
there is a cut-off in the power spectrum on small scales, dependent on
$\Omega_Mh^2$. When this scale is observed in a low-redshift galaxy
power spectrum, its position is dependent on $\Omega_Mh$, where
$h=H_0/100\,{\rm km}\,{\rm s}^{-1}\,{\rm Mpc}^{-1}$, as projection
effect introduce another factor of $h$.

\subsection{Neutrino masses}

The same principal of gravitational collapse versus pressure support
can be applied in the case of massive neutrinos. Initially the
neutrinos are relativistic and their Jeans scale grows with the
horizon. As their temperature decreases their momenta drop, they
become non-relativistic, and the Jeans scale decreases -- they can
subsequently fall into perturbations. Massive neutrinos are
interesting because even at low redshifts the Jeans scale is
cosmologically relevant. Consequently the linear power spectrum (the
fluctuation distribution excluding the non-linear collapse of
perturbations) is not frozen shortly after matter-radiation
equality. Instead its form is still changing at low
redshifts. Additionally, the growth rate depends on the scale - it is
suppressed until neutrinos collapse into perturbations, simply because
the perturbations have lower amplitude.

Both the imprint of the matter-radiation equality scale and neutrino
masses are further complicated by galaxy bias (see
Section~\ref{sec:bias}), which limits measurements made from galaxy
surveys. However, cosmological neutrino mass measurements are still of
strong interest as the standard model of particle physics links
together cosmological photon and neutrino species densities: based on
current photon density (as measured from the CMB), we expect a
cosmological neutrino background with a density $112$\,cm$^{-3}$ per
species, which leads to an expected cosmological density
\begin{equation}
  f_\nu=\frac{\Omega_\nu}{\Omega_M}
    =\frac{\sum m_\nu}{93\Omega_Mh^2\,{\rm eV}}\,.
\end{equation}
Thus a measurement of the cosmological density directly gives a
measurement of the summed neutrino mass, and has the potential to
provide information on the mass hierarchy: Neutrinos come in three
flavours, and detectors measure differences among the three mixed
states, but the masses themselves are unknown. An interesting question
is whether the biggest difference marks the heaviest mixed state (the
so-called normal hierarchy) or the lightest one (an inverted
hierarchy).

\subsection{Baryons}  \label{sec:baryons}

At early epochs baryons are coupled to the photons and are subject to
radiation pressure. If we consider a single fluctuation existing after
inflation, then a spherical shell of baryonic material and photons is
driven away from the perturbation by this pressure. When the photons
and gas decouple, a spherical shell of baryons is left around a
central concentration of dark matter. As the perturbation evolves
through gravity, the density profiles of the baryons and dark matter
grow together, and the final perturbation profile is left with a small
increase in density in a spherical shell at a radial location
corresponding to the sound horizon at the end of the Compton drag
epoch $r_d$: this is the radius of the spherical shell
\shortcite{bashinsky1,bashinsky2}. In order to understand the effect
of this process on a field of perturbations, one can imagine many of
these superimposed ``waves'' propagating simultaneously, resulting in
a slight preference for perturbations separated by the scale of the
sound horizon (perturbations at the original location, and at the
spherical shell). When translated into the power spectrum, this
becomes a series of oscillations in the same way that the Fourier
transform of a top-hat function is a Sinc function.

In addition to these features (called Baryon Acoustic Oscillations, or
BAO), fluctuations smaller than the Jeans scale, which tracks the
sound horizon until decoupling, do not grow, while large fluctuations
are unaffected and continue to grow. The presence of baryons therefore
also leads to a reduction in the amplitude of small scale
fluctuations. A description of the physics for baryons is given by
\shortciteN{Eis98} or Appendix A of \shortciteN{MeiWhiPea}, and a
discussion of the acoustic signal in configuration space can be found
in \shortciteN{EisSeoWhi07}.

In an evolved density field at low redshifts, BAO are damped on small
scales due to large-scale bulk flows, which are well described as
being random \shortcite{Eis07a}. If we write the original power
spectrum as $P_{\rm lin}(k)$, and a version without BAO as $P_{\rm
  nw}(k)$, then
\begin{equation}
  P(k) = P_{\rm lin}(k) e^{-\frac{k^2\sigma^2}{2}}+P_{\rm
    nw}(k)\left(1-e^{-\frac{k^2\sigma^2}{2}}\right)\,,
\end{equation}
where $\sigma$ controls the amplitude of the damping. The observed
damping is expected to be stronger along the line-of-sight due to the
contribution from Redshift-Space Distortions (see
Section~\ref{sec:RSD}).

\section{Physical Processes}

We saw in the previous section that the large-scale distribution of
perturbations in the matter field has the potential to provide a
wealth of cosmological information, as it encodes a number of
different physical processes. When we observe the field by means of a
galaxy survey, we see the evolved product of the linear field, as
traced by galaxies. The differences between the galaxy field and the
original linear matter perturbations can be understood using a number
of simple models that help to explain how the build-up of structure
leads to galaxy formation.

\subsection{Linear structure growth}  \label{sec:lin_growth}

The simplest model for cosmological structure growth follows from
Birkhoff's theorem, which states that a spherically symmetric
gravitational field in empty space is static and is always described
by the Schwarzchild metric \shortcite{birkhoff23}. Thus, a homogeneous
sphere of uniform density with radius $a_p$ (subscript $p$ refers to
the perturbation) can itself be modelled using the same equations that
govern the evolution of the Universe, with scale factor $a$. From this
we can derive the linear growth rate (e.g. \shortciteNP{peebles_lss}).
We assume that we have two spheres containing the same mass, one of
background density $\rho$, and one with density $\rho_p$, where
\begin{equation}
   \rho_p a_p^3=\rho a^3, \,\,\,\delta\equiv\rho_p/\rho-1\,,
\end{equation}
giving, to first order in $\delta$, $a_p=a(1-\delta/3)$. The
cosmological equation for both the spherical perturbation and the
background is 
\begin{equation} 
  \frac{1}{a}\frac{d^2a}{dt^2}=
    -H_0^2\left[\frac{1}{2}\Omega_Ma^{-3}-\Omega_\Lambda\right]\,,
    \label{eq:cosmo}
\end{equation}
where $a$ should be replaced by $a_p$ in the matter density term for
the perturbation, and we assume a $\Lambda$CDM background
model. Using this equation for both spheres, and substituting in the
linear equation for $\delta$ as a function of $a$ and $a_p$ gives, to
first order
\begin{equation} 
  \frac{3}{2}\Omega_M(a) = \frac{d^2\ln\delta}{d\ln a^2} 
    + \left(\frac{d\ln\delta}{d\ln a}\right)^2
    + \frac{d\ln\delta}{d\ln a}
    \left\{1-\frac{1}{2}\Omega_M(a)-\Omega_\Lambda\right\}\,,
\end{equation}
where we have simplified this equation by including the evolution of
the matter and $\Lambda$ densities with respect to the critical density,
\begin{equation}
  \Omega_M(a)=\frac{\Omega_Ma^{-3}}{E^2(a)},\,\,\,\,
  \Omega_\Lambda(a)=\frac{\Omega_\Lambda}{E^2(a)}\,.
\end{equation}
$E(a)$ was given in Eq.~\ref{eq:Frie}.  This simple derivation, and
the solutions to this equation $\delta(z)\propto G(z)$, where $G(z)$
is the linear growth rate, show how large-scale densities grow within
the background Universe.  More details of this derivation, and a
formula valid for general non-interacting Dark Energy models, are
given in \shortciteN{percival05}.

\subsection{Spherical collapse}  \label{sec:sph_coll}

In the above subsection, we saw how the evolution of spherical
perturbations follows that of {\it mini-Universes}, with different
initial densities. A first-order solution was presented, giving the
linear growth rate $G(z)$. Going beyond the first order, the evolution
of the perturbations can be followed further, until the effects of the
assumption of homogeneity become apparent. Closed cosmological models,
which are analogous to collapsed objects, start with sufficient
density that the gravitational attraction overcomes Dark Energy
leading to eventual collapse. Open cosmological models, which are
analogous to under-dense regions, start with so little density that
gravity cannot influence the initial expansion.

\begin{figure}
  \begin{center}
  \includegraphics[width=0.9\textwidth]{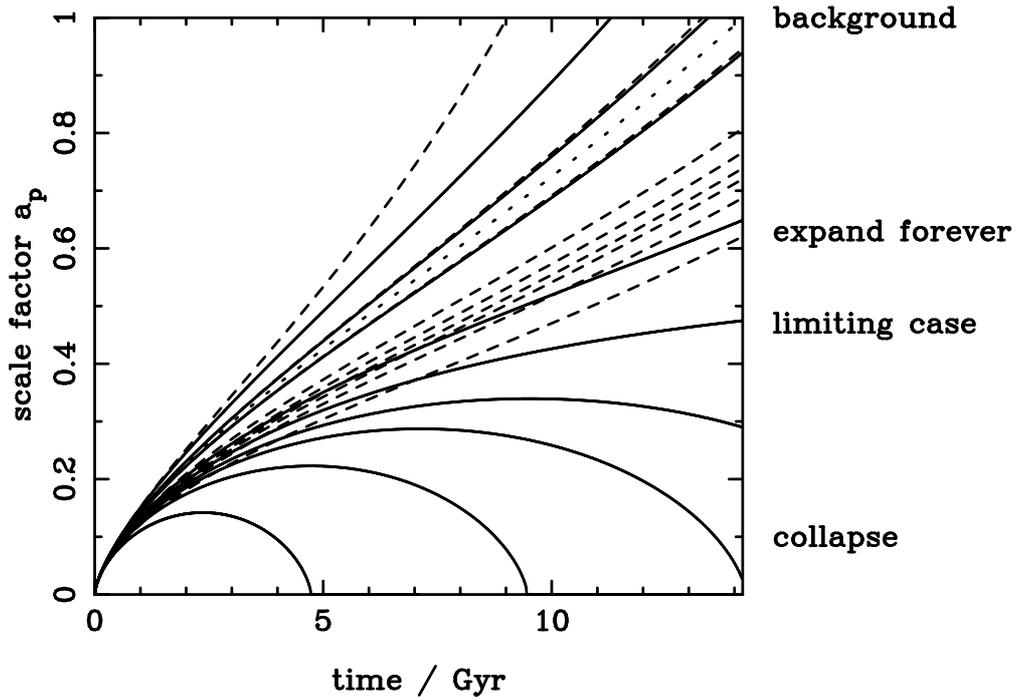}
  \end{center}
  \caption{The evolution of the perturbation scale factor $a_p$, as a
    function of the local density of the patch of universe under
    consideration (solid lines). These were calculated for a flat
    $\Lambda$CDM cosmology, with $\Omega_M=0.25$. The dashed lines
    show the linear extrapolation of the initial perturbation
    over-density. The dotted line shows the background
    expansion. Within the simple spherical collapse model,
    sufficiently over-dense regions collapse to singularities, while
    under-dense regions expand forever. The relationship between the
    linear growth of perturbations (discussed in
    Section~\ref{sec:lin_growth}), and the non-linear evolution
    (discussed in Section~\ref{sec:sph_coll}) can clearly be seen by
    comparing solid and dashed lines.}
  \label{fig:aevol}
\end{figure}

Fig.~\ref{fig:aevol} shows the evolution of the perturbation scale for
various initial conditions, and the link to the linear growth of
perturbation scale. Given a low initial density, such as expected for
a ``void'', the perturbation will expand more rapidly than the
background, whereas more dense regions, such as large groups of
galaxies, collapse. Within this simple model, collapse to a
singularity is predicted; in practice the granulation of material
leads to virialisation at a fixed radius.

At any epoch, there is a critical initial density for collapse:
perturbations that are more dense have collapsed by the epoch of
interest, while less dense perturbations have not. For ease, the
critical density is usually extrapolated using linear theory to
present-day, and is called the linearly-extrapolated critical density
for collapse $\delta_c$. $\delta_c$ has a weak dependence on
cosmological model, and a review of methods to calculate $\delta_c$ is
given in, for example, \shortciteN{percival05}. Given the weak
dependence it is commonly approximated by its value for an Einstein-de
Sitter cosmological model
\begin{equation}
  \delta_c \simeq 1.686 \frac{G(0)}{G(z)}\,,
  \label{eq:dc_approx}
\end{equation}
where $G(z)$ is the standard linear growth rate, which extrapolates the
initial over-density to present day.

\subsection{Press-Schechter theory}

Taking the spherical model one step further it is possible to form a
``complete'', although simplistic, model for structure growth as a
function of the mass of an object. \shortciteN{ps} introduced the
concept of smoothing the initial density field to determine the
relative abundances of perturbations with different mass. For a filter
of width $R$, corresponding to a mass scale $M=4/3\pi \rho R^3$ where
$\rho$ is the average density of the Universe, smoothing the matter
field with power spectrum $P(k)$ gives a further field with spatial
variance
\begin{equation}
  \sigma_M^2=\frac{1}{2\pi}\int^\infty_0P(k)W^2(k,R)k^2\,dk\,,
\end{equation}
where $W(k,R)$ is the window function for the smoothing. If a sharp
k-space filter is used then, around any spatial location, the smoothed
over-density forms a Brownian random walk in $\delta$, as a function of
filter radius (usually plotted as $\sigma_M^2$). Comparing the
relative numbers of trajectories that have crossed the critical
density (as discussed in the previous section) at a mass greater than
that of interest (some may cross at high mass, but be below the
critical density at low mass, but still be counted as collapsed
objects), gives a distribution of haloes with mass (commonly called a
``mass function'')
\begin{equation}
  f(>M)=2\int^\infty_{\delta_c}\frac{\delta_c}{\sqrt{2\pi\sigma_M^2}}
    \exp\left(-\frac{\delta^2}{2\sigma_M^2}\right)\,d\delta\,.
\end{equation}
Thus we have a complete model for structure formation: smoothing the
fluctuations leads to the masses of collapsed objects, while the
spherical perturbation model gives the epoch of collapse for those
perturbations that are sufficiently dense. Comparing with numerical
simulations reveals that, while this is a good model, it cannot
accurately predict the measured redshift-dependent mass function, and
fitting formulae to N-body simulations are the currently preferred way
to model this (e.g. \shortciteNP{sheth99,jenkins01}).
 
\subsection{Galaxy bias}  \label{sec:bias}

The above discussion only considered concentrations of mass, which are
commonly referred to as ``haloes'', whereas the most easily observed
extragalactic objects are galaxies. In general, a sample of galaxies
will form a Poisson sampling of some underlying field with over-density
$\delta_g$, but this field may be different from the matter field,
which has over-density $\delta_M$. Here, a subscript $g$ refers to
galaxies, and $M$ to the matter. This ``bias'' between fields encodes
the physics of galaxy formation, and may therefore be a general,
non-local, non-linear function. On large-scales we will now see that this
bias reduces to a local, linear function $b\equiv\delta_g/\delta_M$.

The simplest model for large-scale galaxy bias is the peak-background
split model \shortcite{BBKS,cole89}. Here it is postulated that halo
formation depends on the local density field $\delta_M$. Large scale
density modes $\delta_l$ can alter the local halo density $n$ by
pushing pieces of the density field above a critical threshold,
leading to a change in the number density of haloes,
\begin{equation}
  n\to n-\frac{dn}{d\delta_c}\delta_l\,.
\end{equation}
To first order, the addition of the long wavelength mode leads to a
bias, 
\begin{equation}
  b\equiv\frac{\delta_{\rm new}}{\delta_l}=1-\frac{d\ln n}{d\delta_c}\,,
\end{equation}
where $\delta_{\rm new}$ is the boosted over-density.  Thus we see how
the shape of the mass function gives the large-scale (linear) bias of
haloes. On large scales we expect a linear relationship between the
large scale mode amplitudes and the change in number density, so the
shape of the galaxy and mass power spectra are the same.

\begin{figure}
  \begin{center}
  \includegraphics[width=0.7\textwidth]{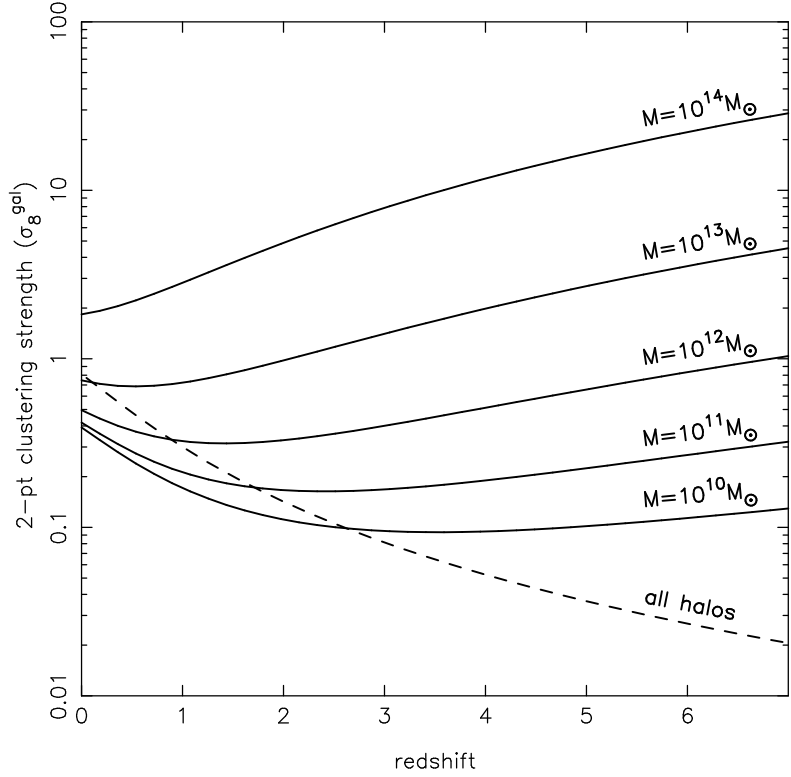}
  \end{center}
  \caption{The expected large-scale clustering amplitude for haloes of
    different mass as a function of scale, compared to the linear
    growth rate. Clustering has been estimated using the
    peak-background split model for a standard $\Lambda$CDM
    cosmological model. Haloes are expected to grow in mass with time,
    so will not evolve along the lines shown, which link objects of
    the same mass at different times. The dashed line shows the
    expected growth in the clustering of all of the mass, as given by
    $G(z)$.}
  \label{fig:clustering_strength}
\end{figure}

Fig.~\ref{fig:clustering_strength} shows the result of the
peak-background split model, giving the amplitude of clustering for
haloes of different mass as a function of redshift. Higher mass haloes
have stronger clustering as a result of the steeper mass function,
which means that a small long-wavelength perturbation can have a
disproportionate effect on the number density of haloes.

On small scales, the peak-background split model no longer holds, and
we must look to other models for the clustering of galaxies. One such
model is the ``halo'' model \shortcite{seljak00,peacock00,cooray02}, which
is a simple framework where galaxies trace halos on large scales, with
bias determined by the distribution of halo masses within which the
galaxies reside. On small scales, we expect galaxy clustering to be
different from that of the halos, as pairs of galaxies inside single
collapsed structures become important. The halo model simply combines
these two scenarios into a full-scale model for galaxies.

\section{Galaxy Survey basics}

In this section we explore some of the practical ideas and methods
used to analyse a galaxy survey and make cosmological
measurements. The clustering signal is a 3-dimensional signal, so
ideally we wish to know the distribution of matter (or galaxies) in
3D. However, while the angular positions of galaxies are in general
easy to measure, the radial distances are not so easy. In order to
estimate distances to galaxies, large surveys typically use the
spectrum of light emitted by galaxies, including absorption and
emission lines and more general features, all resulting from the
integrated stellar light. By comparing these features to rest-frame
models, the redshifts of the galaxies can be estimated and, using the
Hubble expansion rate, their distances.

It is possible to fit observed broad-band colours with templates or
training samples of emitted light profiles, and thus estimate galaxy
recession velocities. These ``photometric redshifts'' - redshifts
estimated from broad-band colours only - vary in quality between
galaxy samples. For example, they can have offsets from the true
redshifts with standard deviations of $\sim0.03(1+z)$ for red galaxies
with strong 4000\,\AA\ breaks \shortcite{ross11}, while more general
populations can give estimates with $\sigma_z\sim0.05(1+z)$. The level
of precision depends on the photometric quality of the data used, the
number of bands, etc. If spectra can be obtained for galaxies, then
absorption and emission lines can be used to estimate redshifts with
typical errors of $\sim0.001...0.0001(1+z)$, allowing clustering along
the line-of-sight to be more accurately measured.

\subsection{Overview of galaxy surveys}

Recent advances in galaxy surveys have largely been driven by new
instrumentation that enables multiple galaxy spectra to be obtained
simultaneously. Using these multi-object spectrographs (MOS), a number
of survey teams have created maps of many hundreds of thousands or
millions of galaxies. Key wide-field facilities include the AA$\Omega$
instrument on the Anglo-Australian Telescope, which has been used to
conduct the 2-degree Field Galaxy Redshift Survey (2dFGRS;
\shortciteNP{colless03}) and Wigglez \shortcite{drinkwater10} surveys,
and the Sloan Telescope, which has been used for the Sloan Digital Sky
Survey (SDSS; \shortciteNP{york00}).

In following sections we will use results from the Baryon Oscillation
Spectroscopic Survey (BOSS \shortciteNP{dawson13}), part of the
SDSS-III project \shortcite{eisenstein11} to demonstrate
possible measurements. BOSS is primarily a spectroscopic survey, which is
designed to obtain spectra and derive redshifts from them for 1.2
million galaxies over an extragalactic footprint covering $10\,000$
square degrees.  $1000$ spectra are simultaneously recovered in each
observation, using aluminium plates with drilled holes to locate
manually plugged optical fibres that feed a pair of double
spectrographs. Each observation is performed in a series of
$900$-second exposures, integrating until a minimum signal-to-noise
ratio is achieved for the faint galaxy targets.  The resulting data
set has high redshift completeness $>97$ per cent over the full survey
footprint.

\subsection{Measuring over-densities}

As described in Section~\ref{sec:delta}, we need to translate from an
observed galaxy density to a dimensionless over-density. For a galaxy
survey, this means understanding where you could have observed
galaxies (called the survey {\em mask}), not just where galaxies were
observed. For current surveys, the mask can be split into independent
radial and angular components, where the angular component of the mask
includes the varying completeness and purity of observations and the
radial component depends on the galaxy selection criteria. For a
magnitude-limited catalogue, the radial distribution can be determined
by integrating under a model luminosity function \shortcite{cole11},
while for a colour selected sample, a fit to the galaxy distribution
is usually performed \shortcite{And12}. The mask is commonly
quantified by means of a random catalogue - a catalogue of spatial
positions that Poisson sample the expected density $\bar{\rho}$ (see
Eq.~\ref{eq:delta}), as outlined by the mask. This catalogue has the
same spatial selection function as the galaxies but no clustering.

If the correct mask has been used, the estimate of $\delta_g({\bf x})$
is unbiased, but has spatially varying noise, and we only have
estimates within the volume covered by the survey. In order to
optimally measure the amplitude of clustering, we need to apply
weights. If we assume that the galaxy field forms a Poisson sampling
of a field with expected power spectrum $\bar{P}(k)$, then the optimal
weights to use are
\begin{equation}  \label{eq:wght_FKP}
   w({\bf x})=\frac{1}{1+\bar{n}({\bf x})\bar{P}(k)}\,,
\end{equation}
where $\bar{n}({\bf x})$ is the expected density of galaxies
\shortcite{FKP}. The dependence on both $\bar{n}({\bf x})$ and
$\bar{P}(k)$ in these weights balances the shot-noise and
sample-variance components of the measurement error. For multiple
samples of galaxies covering the same volume, but with different
expected clustering strengths (each sample is given a subscript $i$),
the optimal weights \shortcite{PVP} to apply are
\begin{equation}
   w_i({\bf x})=\frac{\bar{P}_i(k)}{1+\sum_i\bar{n}_i({\bf x})\bar{P}_i(k)}\,.
\end{equation}
These weights provide the same balance between shot-noise and sample
variance as Eq.~\ref{eq:wght_FKP}, but now also balance this against
the clustering strength of each sample. For example, galaxies with
high clustering strength are up-weighted as these contain higher
signal-to-noise. These weighting schemes assume that galaxies Poisson
sample the underlying matter field, but this is only an
approximation. In fact, galaxies live in matter haloes, and haloes do
not Poisson sample the mass in the Universe: on large-scales the
matter has a distribution with sub-Poisson variance. Thus applying a
weighting based on the mass of the haloes within which the galaxies
reside can reduce the shot noise in surveys \shortcite{seljak09}.

\subsection{Measuring the power spectrum}  \label{sec:pk_est}

If we only wish to measure the isotropically averaged power spectrum,
then it is possible to simply measure this directly from  a Fourier
transform of the over-density field \shortcite{FKP}. Suppose that we
have a galaxy catalogue with density $n_g({\bf x})$, and a random
catalogue describing the mask with density $n_r({\bf x})$ containing
$\alpha$ times as many objects. Following \shortciteN{FKP}, we start
from the un-normalised over-density field
\begin{equation}
  F({\bf x}) = n_g({\bf x}) - n_r({\bf x})/\alpha\,.
\end{equation}
Taking the Fourier transform of this field, and calculating the power
gives
\begin{equation}  \label{eq:fkp_pk}
  \langle |F({\bf k})|^2 \rangle = \int \frac{d^3k'}{(2\pi)^3} 
    [P({\bf k'})-P(0)\delta_D({\bf k})] |G({\bf k}-{\bf k'})|^2
   + (1+\frac{1}{\alpha})\int d^3x \bar{n}({\bf x})\,,
\end{equation}
where $G({\bf k})$ is the Fourier transform of the window function,
defined by 
\begin{equation}
  G({\bf k}) \equiv \int\bar{n}({\bf x})e^{i{\bf k.r}}d^3x\,,
\end{equation} 
and the final term in Eq.~\ref{eq:fkp_pk} gives the shot noise. We see
that the effect of the survey mask, which is multiplicative in
real-space (think of seeing the Universe through a window only
revealing the patch observed), becomes a convolution in
Fourier-space. There is a shot noise contribution on all scales, which
can be subtracted from the power spectrum estimates, although it still
contributes to the error. The fact that we have to estimate the mean
density of galaxies from the sample itself means that the $k=0$ mode
is artificially set to zero, equivalent to subtracting a single Dirac
delta function from the centre of the un-convolved power. The
equivalent correction for correlation function estimates is commonly
know as the ``Integral constraint''

\begin{figure}
  \begin{center}
  \includegraphics[width=0.7\textwidth]{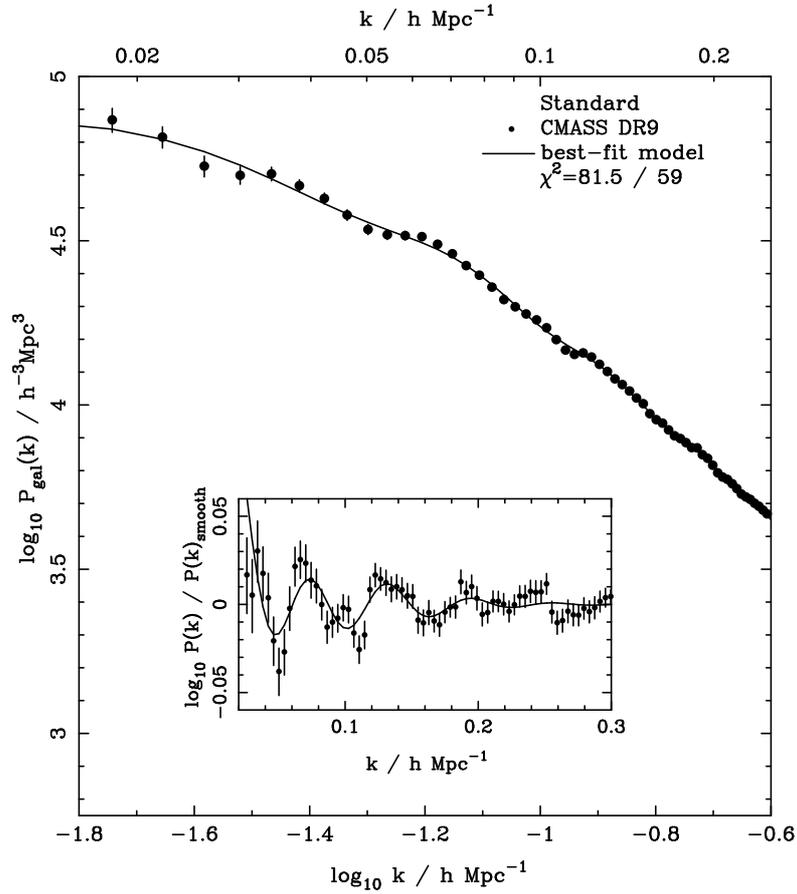}
  \end{center}
  \caption{The spherically averaged BOSS DR9 power spectrum for the
    CMASS galaxy sample (solid circles with 1$\sigma$ errors),
    calculated using the method of \protect\shortciteN{FKP}. The
    vertical dotted lines show the range of scales fitted
    ($0.02<k<0.3$\,h\,Mpc$^{-1}$), and the inset shows the BAO within
    this k-range, determined by dividing both model and data by the
    smooth best-fit model. Plot from \protect\shortciteN{And12}.}
  \label{fig:BOSS_DR9_pk}
\end{figure}

The comoving power spectrum, as measured from the Data Release 9 (DR9)
BOSS data, using the \shortciteN{FKP} technique is shown in
Fig.~\ref{fig:BOSS_DR9_pk}. The inset shows the ratio of the measured
power to a smooth fit, isolating the signature of baryons (see
Section~\ref{sec:baryons}), which is clearly visible
\shortcite{And12}.

We cannot easily measure the anisotropic power spectrum as a function
of $\mu$ (see Section~\ref{sec:aniso}) using Fourier methods because
the line-of-sight varies across a survey, and does not line-up with a
Cartesian grid. Thus, to measure the anisotropic power spectrum, one
either needs to manually perform the transform, so a different
line-of-sight can be used for each galaxy pair \shortcite{yamamoto},
or decompose into a basis that is itself separable along and across
the spatially varying line-of-sight (e.g. \shortciteNP{heavens95}). As
surveys increase in size and statistical errors reduce, methods such
as these will become increasingly important.

\subsection{Measuring the correlation function}  \label{sec:xi_est}

In order to estimate the correlation function, we can directly make
use of the idea of ``throwing down sticks'' described in
Section~\ref{sec:xi}. Suppose that the mask is quantified by a random
catalogue as in Section~\ref{sec:pk_est}, then we can form an estimate
of the correlation function $\bar{\xi}$
\begin{equation}  \label{eq:xi_obs}
  \bar{\xi}(r) = \frac{DD(r)}{RR(r)}-1\,,
\end{equation}
where $DD(r)$ is the number of galaxy-galaxy pairs within a bin with
centre $r$ normalised to the maximum possible number of galaxy-galaxy
pairs (ie. for $n$ galaxies the maximum number of distinct pairs is
$n(n-1)/2$). $RR(r)$ is the normalised number of random-random pairs,
and we can similarly define $DR(r)$ as the normalised number of
galaxy-random pairs.

This estimate of $\xi(r)$ is biased and we find that
\begin{equation}  \label{eq:xi_expect}
  \langle 1+\bar{\xi}(r) \rangle =
  \frac{1+\xi(r)}{1+\xi_\Omega(r)}\,,
\end{equation}
where $\xi_\Omega(r)$ is the mean of the two-point correlation
function over the mask \shortcite{landy93}: this corrects for the fact
that the true mean density of galaxies $\bar{n}({\bf r})$ is estimated
from the sample itself, normalising the total number of pairs. For
small samples, the factor $[1+\xi_\Omega(r)]$, called the ``integral
constraint'', limits the information that can be extracted from
$\bar{\xi}(r)$ - in the limit where we only observe pairs of a single
separation, $\xi_\Omega\simeq\xi$, and $\bar{\xi}(r)$ contains no
information.

Because the galaxy and random catalogues are uncorrelated, $\langle
DR(r)\rangle=\langle RR(r)\rangle$, and we can consider a number of
alternatives to Eq.~\ref{eq:xi_obs}. In particular
\begin{equation}
  \bar{\xi}(r) = \frac{DD(r)-2DR(r)+RR(r)}{RR(r)}\,,
\end{equation}
has been shown to have good statistical properties
\shortcite{landy93}, although Eq.~\ref{eq:xi_expect} still holds for
this estimator (ie. it still ``suffers'' from the integral
constraint). Pair counting methods can be trivially extended to bin
pairs in separation and angle to the line-of-sight, giving estimates
of the anisotropic correlation function. It is also possible to
directly integrate over $\mu$, weighted by a Legendre polynomial to
give estimates of the multipoles, or by top-hat windows in $\mu$
(called ``Wedges''; \shortciteNP{Kaz12}). These moments of the
correlation function provide a mechanism to compress the anisotropic
correlation function while still retaining the majority of the
available information.

\begin{figure}
  \begin{center}
  \includegraphics[width=0.7\textwidth]{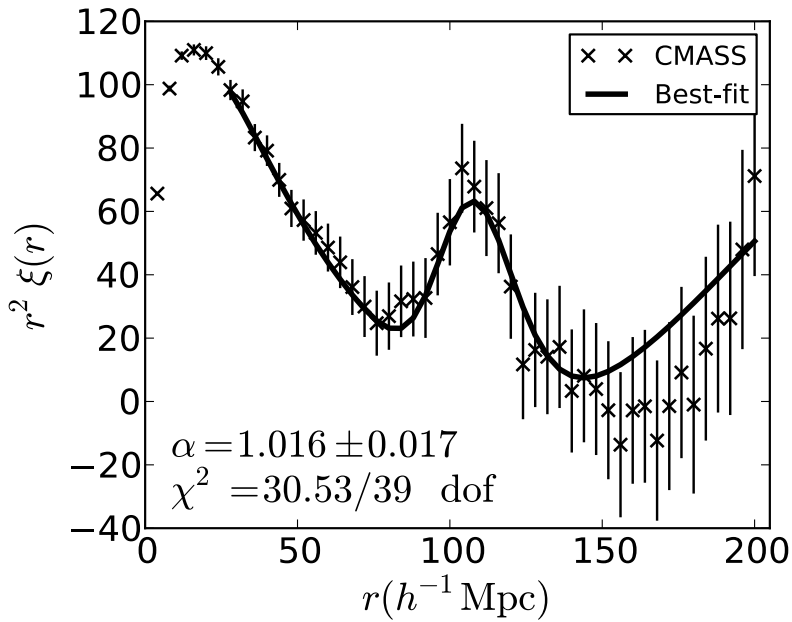}
  \end{center}
  \caption{The spherically averaged BOSS DR9 correlation function for
    the CMASS galaxy sample (solid circles with 1$\sigma$ errors),
    calculated using the method of \protect\shortcite{landy93}. Plot
    from \protect\shortciteN{And12}.}
  \label{fig:BOSS_DR9_xi}
\end{figure}

An example correlation function, calculated from the BOSS DR9 CMASS
sample \shortcite{And12} is given in Fig.~\ref{fig:BOSS_DR9_xi}. The
BAO ``bump'' caused by the physics described in
Section~\ref{sec:baryons} is clearly visible. On large scales, the
correlation function reduces in amplitude (note that $r^2\xi$ is
plotted, enhancing the appearance of the large-scale noise), showing
that the galaxies are evenly distributed. In fact, conservation of
pair number means that the correlation function changes sign on very
large scales.

\subsection{Reconstructing the linear density}

In an evolved density field, pairs of over-densities initially
separated by the BAO scale are subject to bulk motions that ``smear''
the initial pair separations. The matter flows and peculiar velocities act on
intermediate scales ($\sim20$\,Mpc\,h$^{-1}$), and therefore suppress
small-scale oscillations in the matter power spectrum and smooth the
BAO feature in the correlation function
\shortcite{EisSeoWhi07,CroSco08,Mat08a,Mat08b}. As galaxies trace the
matter density field, the BAO in galaxy surveys are also
damped. \shortciteN{Eis07a} suggested that this smoothing can be
reversed, in effect using the phase information within the density
field to reconstruct linear behaviour. Although not a new idea
(e.g. \shortciteNP{Pee89,Pee90,Nuss92,Gram93}), the dramatic effect on
BAO recovery had not been previously realised, and the majority of the
benefit was shown to be recovered from a simple reconstruction
prescription.

\begin{figure}
  \begin{center}
  \includegraphics[width=0.7\textwidth]{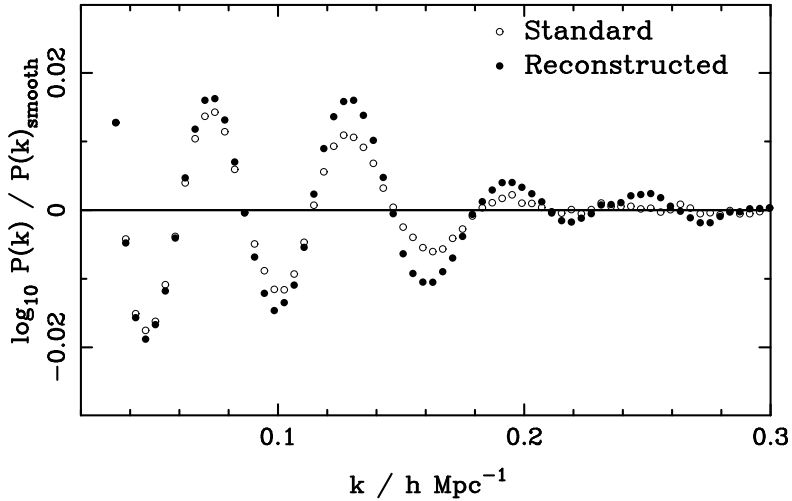}
  \end{center}
  \caption{The ratio between recovered power spectrum and smooth fit
    for the 600 BOSS DR9 CMASS catalogues, before and after applying a
    simple reconstruction algorithm. We see that, on average,
    reconstruction acts to improve the BAO signal on small
    scales. Plot from \protect\shortciteN{And12}.}
  \label{fig:mock_recon}
\end{figure}

``Reconstruction'' has been used to sharpen the BAO feature and
improve distance constraints on mock data
\shortcite{PadWhi09,Noh09,Seo10,Meh11}, and was recently applied to
the SDSS-II Luminous Red Galaxy (LRG) sample \shortcite{Pad12}. The
reconstruction was particularly effective in this case, providing a
1.9 per cent distance measurement at $z=0.35$, decreasing the error by
a factor of 1.7 compared with the pre-reconstruction measurement. On
the BOSS DR9 CMASS sample, however, reconstruction was not as
effective, and did not significantly reduce the error on the BAO scale
measurement. \shortciteN{And12} showed, using mock catalogues, that
this is expected, and that the effectiveness of reconstruction varies
across mocks. The average effect of reconstruction on the BOSS DR9
CMASS mocks is shown in Fig.~\ref{fig:mock_recon}.

\subsection{Ly-$\alpha$ forest surveys}

Light from distant quasars passes through intervening clouds of gas
that absorb ultraviolet light predominantly at the wavelength of the
Lyman alpha transition in neutral hydrogen (122\,nm). The absorbing
clouds are all at lower redshift than the quasar, so absorption lines
are on the shorter wavelength side of the quasar emission line. Thus,
if we divide the quasar spectrum by the expected continuum for each
quasar, we are left with a 1D absorber over-density profile, which can
be used to estimate the matter over-density field along the
line-of-sight in much the same way that galaxies can be used directly
(in fact the link between absorber density and matter density is more
complicated than between galaxies and the mass). The great thing about
using the Ly-$\alpha$ forest is that a single spectrum gives
information about multiple structures along the line of sight, and
thus this method provides a highly efficient use of telescopes. Using
quasars with $z>2.15$, the BOSS has made BAO measurements using Lyman
$\alpha$ forest in quasar spectra, extending BAO measurements into the
redshift range $2<z<3.5$ \shortcite{Busca13,Slosar13,Kirkby13}.

\section{Observational effects}  \label{sec:aniso_effects}

Up to this point, we have been concerned with comoving galaxy
clustering, and the information it contains. In fact, a lot of the
information from galaxy surveys does not come directly from the
comoving power, but from effects that distort the observed signal away
from this ideal. We now outline three of these effects, showing how
they can be used to retrieve cosmological information.

\subsection{Projection and the Alcock-Paczynski
  effect}  \label{sec:AP}

The physics described above is encoded into the comoving galaxy power
spectrum. However, we do not measure clustering directly in comoving
space, but we instead measure galaxy redshifts and angles and infer
distances from these. If we use the wrong cosmological model to do
this conversion, then the distances we infer will be wrong and the
comoving clustering will contain detectable distortions.

In the radial direction, provided that the clustering signal is small
compared with the cosmological distortions, we are sensitive to the
Hubble parameter through $1/H(z)$. In the angular direction the
distortions depend on the angular diameter distance
$D_A(z)$. Adjusting the cosmological model to ensure that angular and
radial clustering match constrains $H(z)D_A(z)$, and was first
proposed as a cosmological test (the AP test) by \shortciteN{AP}. If
we instead consider averaging clustering in 3D over all directions,
then, to first order, matching the scale of clustering measurements to
the comoving clustering expected is sensitive to
\begin{equation}
  D_V(z)=\left[(1+z)^2D_A^2(z)\frac{cz}{H(z)}\right]^{1/3}\,.
\end{equation}

Although this projection applies to all of the clustering signal, BAO
provide the most robust and strongest source for the comparison
between observed and expected clustering, providing a distinct feature
on sufficiently large scales that it is difficult to alter with
non-linear physics. In order to extract this information, we need to
parametrize over nuisance broad-band features, extracting just the BAO
signal. This is usually achieved by means of a smooth function,
commonly a polynomial or spline with a set of free parameters, with
sufficient flexibility to match the broad-band shape of all of the
cosmological models to be tested, but not the BAO signal. If a
fiducial cosmological model is used to convert redshifts to distances,
then departures between the expected BAO position, and the observed
one are commonly quantified by dilation scales. For an anisotropic
fit, we can define two dilation scales, perpendicular and parallel to
the line-of-sight
\begin{equation}
  \alpha_{\perp} = \frac{D_{A}(z)r_d^{\rm fid}}{D^{\rm fid}_{A} r_d} \,,\,\,\,\,
  \alpha_{\parallel} = \frac{H^{\rm fid}(z)r_d^{\rm fid}}{H(z) r_d} \,,
\end{equation}
where $r_d$ sets the comoving BAO scale (see
Section~\ref{sec:baryons}), and a superscript ${\rm fid}$ denotes that
a quantity refers to the fiducial cosmology used to measure the
clustering signal. For a single fit to the BAO in an isotropically
averaged clustering measurement, the combined dilation scale is
\begin{equation}
  \alpha = \frac{D_V(z)r_d^{\rm fid}}{D^{\rm fid}_Vr_d} \,.
\end{equation}

Estimates of the BAO scale can be made based on either $\xi(r)$ or
$P(k)$, but they include the noise from small scales and shot noise
differently. \shortciteN{And12} averaged measurements made from both
statistics, using mocks to show that this improved the combined error.

A review of current BAO measurements was provided in the introduction
of \shortciteN{And12}, which described recent experiments
(e.g. \shortciteNP{Beutler11,Bla11a,Pad12}), leading to the analyses
presented in \shortcite{And12}. Since then, the DR9 data set has been
analysed splitting into measurements along and across the
line-of-sight \shortcite{And13}, and the next release of measurements
from BOSS is imminent. In addition, the range of redshifts covered by
BAO measurements has increased further, through BOSS analyses using
the Lyman-$\alpha$ forest in quasar spectra
\shortcite{Busca13,Slosar13,Kirkby13}.

\subsection{Redshift-space distortions}  \label{sec:RSD}

The second observational effect that we will consider results from the
fact that our estimates of galaxy distances are made from
redshifts. These redshifts are caused by both the Hubble expansion and
from any additional motion within a comoving frame, called the
peculiar velocity. We can write
\begin{equation}
  {\bf s}({\bf r}) = {\bf r}-v_r(r)\frac{\bf r}{r}\,,
\end{equation}
where ${\bf s}$ is the redshift-space (i.e. derived from the redshift)
position of a galaxy, ${\bf r}$ is the true real-space position of the
galaxy away with the observer at the origin, and $v_r(r)$ is the
radial component of the peculiar velocity. The distortions in the
field that result from the peculiar-velocity dependent shifts are
called Redshift-Space Distortions (RSD).

Locally, galaxies act as test particles in the flow of matter. If
galaxies fully sampled the velocity field, the peculiar velocity
distribution of galaxies would match that of the mass. We should be
aware that galaxies are expected to be at special locations in the
density field, potentially giving rise to a small bias in their
velocity distribution compared to that of the mass. 

\begin{figure}[t]
  \begin{center}
  \includegraphics[width=0.7\textwidth]{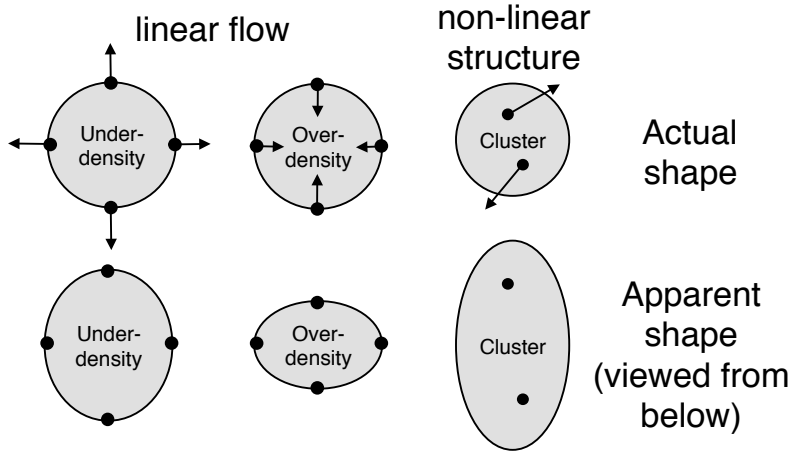}
  \end{center}
  \caption{Explanatory diagram showing how real-space structures (top
    row) look in redshift-space (bottom row).}
  \label{fig:RSD}
\end{figure}

As discussed in Section~\ref{sec:lin_growth}, structures are
continually growing through gravity. On large-scales this growth is
the dominant source of RSD. Consider a galaxy on the near-edge of a
strong over-density: this galaxy will tend to be falling in to the
over-density, away from us, increasing its redshift and moving its
apparent position closer to the centre of the over-density. A galaxy at
the far-edge of an over-density will appear closer, and we therefore
see that clusters will appear ``squashed'' along the line-of-sight in
redshift-space. By a similar argument, under-dense regions appear
``stretched'' along the line-of-sight. This is shown in
Fig.~\ref{fig:RSD}.

If we assume that the patch of the Universe from which clustering is
measured is sufficiently far away that the line-of-sight is
approximately constant across the patch, this effect causes an
increase in the measured power that can be easily modelled. Following
\shortciteN{kaiser87}, on linear scales we can derive a change in the
power corresponding to
\begin{equation}  \label{eq:pk_rsd}
  P_{gg}^s(k,\mu) = P_{gg}^r(k)  + 2\mu^2P_{g\theta}^r(k)
    +\mu^4P_{\theta\theta}^r(k)\,,
\end{equation}
where $\theta=\nabla\cdot{\bf v}$ is the divergence of the peculiar
velocity field, a superscript $s$ denotes that a quantity is measured
in redshift-space and $r$ in real-space, and a subscript $g$ denotes
that a quantity refers to the galaxy field. Furthermore, in the
linear regime, we can write $\theta=-f\delta_m$, where $m$ refers to
the matter field, and
\begin{equation}
  f\equiv\frac{d\ln G}{d\ln a}\,,
\end{equation}
is the logarithmic derivative of the linear growth rate. Assuming a
linear bias, we can write $\delta_g^r=b\delta_m^r$, and we see that
the amplitude of the anisotropic power spectrum given in
Eq.~(\ref{eq:pk_rsd}) depends on $b\sigma_8$ and $f\sigma_8$, where
$\sigma_8$ is the standard deviation of the distribution of
over-densities when averaged over spheres of radius 8\,Mpc\,h$^{-1}$. If
we spherically average the power spectrum, equivalent to integrating
over $0<\mu<1$, we find that
\begin{equation}
   P_{gg}^s = P_{mm}^r\left[b^2+\frac{2}{3}bf+\frac{1}{5}f^2\right]\,.
\end{equation}

On small-scales, there is a second RSD effect caused by galaxy motion
within collapsed objects with deep potential wells. The random
velocities attained by such galaxies smear the collapsed object along
the line of sight in redshift space, leading to the existence of
linear structures pointing towards the observer in
redshift-space. This is shown in Fig.~\ref{fig:RSD}. These apparent
structures are known as ``Fingers-of-God'' (FoG) and reduce
information on small scales. If the pairwise velocity dispersion
within a collapsed object has an exponential distribution
(superposition of Gaussians), then we get a multiplicative damping
term for the power spectrum of the form
\begin{equation}  \label{eq:fog}
  F_{\rm fog}(k,\mu)=\left(1+\frac{k^2\mu^2\sigma_p^2}{2}\right)^{-1}\,,
\end{equation}
where $\sigma_p\sim400\,$kms$^{-1}$ \shortcite{hawkins03}. It is
common to multiply $P_{gg}^s(k,\mu)$ from Eq.~(\ref{eq:pk_rsd}) by
$F_{\rm fog}(k,\mu)$ to provide a combined model for the anisotropic
power spectrum. However this model is not guaranteed to work in the
quasi-linear regime, although the damping term can model some of the
quasi-linear signal if $\sigma_p$ is treated as a free parameter
\shortcite{percival09}.

An alternative to modelling the FoG is to try to correct data by using
phase information to ``collapse clusters'' along the line of sight
(e.g. \shortciteNP{tegmark04}). This method has similarities with
``reconstruction'' used to recover BAO information. For the SDSS-II
Luminous Red Galaxy sample, \shortciteN{reid10} used an anisotropic
friends-of-friends group finder, normalised using numerical
simulations \shortcite{reid09} to locate clusters and move the
galaxies to the average redshift, removing much of the FoG signal and
reducing the modelling burden.

Standard measurements of the RSD signal are intrinsically limited by
the number of radial modes present, which provide the sample variance
limit. \shortciteN{mcdonald09} showed that, with two samples of
galaxies covering the same region, it is possible to measure ratios of
power spectra independent of cosmic variance - provided that the two
samples both have different linear deterministic biases. Taking into
account the anisotropic nature of their redshift-space power spectrum,
shot-noise limited measurements of $f/b_i$ and $b_i/b_j$ (where $i,j$
refer to the galaxy samples) would be possible. These combine to give
measurements of $f\sigma_8$ that are limited by the total number of
modes within a survey, rather than just those in the radial direction
- potentially a strong source of further information.

The picture for RSD measurements presented in this section relies on a
number of assumptions, particularly the combination of linear theory
with a simple damping model, and the assumption that we are only
interested in pairs of galaxies for which the line-of-sights are
approximately parallel. These assumptions have been tested and shown
to be adequate for current RSD growth rate measurements
\shortcite{samushia12}. Many RSD-based measurements of the growth rate
have been made from a number of spectroscopic surveys, with a recent
compilation provided in \shortciteNP{reid12}.

\subsection{Joint AP and RSD measurements}

The primary problem with making RSD and AP measurements of the full
(not just BAO) clustering signal is disentangling both effects
\shortcite{ballinger}, although \shortciteN{samushia12} showed how
future surveys ameliorate this problem. For current surveys, this
separation can only be achieved on large-scales, where RSD can be
modelled with perturbation theory, leading to joint RSD and AP
measurements (e.g. \shortciteNP{reid12}). The power of the large-scale
AP test was shown in \shortciteN{samushia13}, who demonstrated that
using the joint $H(z)$ and $D_A(z)$ AP measurements of
\shortciteN{reid12} led to a fortuitous degeneracy breaking compared
with using the isotropic BAO measurements of \shortciteN{And12}. This
led to a factor of five improvement on measuring the Dark Energy
equation-of-state $w(z=0.57)$, which is particularly impressive given
that the same data were used to make both measurements.

\subsection{Primordial non-Gaussianity}  \label{sec:f_nl}

As discussed in Section~\ref{sec:pk_phys}, determining the amount of
primordial non-Gaussianity in the matter over-density field provides a
key way of distinguishing between inflationary models. Observations of
the large-scale clustering of galaxies have the potential to measure
this, as we will now demonstrate. One of the simplest way to
parameterise non-Gaussianity is to assume that the potential has a
local quadratic term
\begin{equation}  \label{eq:fnl_pot}
  \Phi=\phi+f_{NL}\left(\phi^2-\langle\phi^2\rangle\right)\,.
\end{equation}
Although many other forms of non-Gaussianity are possible, any
detection of non-Gaussianity would be interesting, and it is therefore
instructive to consider how this simple model can be constrained by
galaxy clustering measurements \shortcite{dalal08}. 

We can see this by reconsidering the peak-background split model - in
Section~\ref{sec:bias}, we saw that halo formation is much easier with
additional long-wavelength fluctuations. If we decompose the potential
in Eq.~(\ref{eq:fnl_pot}) into long ($\phi_l$) and short ($\phi_s$)
wavelength components, we find that
\begin{equation}  \label{eq:fnl_pot_ls}
  \Phi=\phi_l+f_{NL}\phi_l^2+(1+2f_{NL}\phi_l)\phi_s
    +f_{NL}\phi_s^2+{\rm cnst}\,.
\end{equation}
The $f_{NL}\phi_l^2$ term is small, but we see that there is an extra
term $2f_{NL}\phi_l\phi_s$, which is of importance for the large-scale
bias. To see how this enters into the peak-background split model,
note that the link between over-density and potential can be written
$\delta_l(k)=\alpha(k)\phi(k)$, where
\begin{equation}
  \alpha(k)=\frac{2c^2k^2T(k)G(z)}{3\Omega_MH_0^2}\,.
\end{equation}
Now, one factor of $\alpha$ gets absorbed into converting one of the
potentials in the extra term to an over-density, but we are left with a
critical density that is modulated not just by the Gaussian
long-wavelength fluctuation, but by 
\begin{equation}
  \delta_l+2f_{NL}\phi_l=\delta_l\left(1+\frac{2f_{NL}}{\alpha(k)}\right)\,.
\end{equation}
This extra term propagates into the change in number density of haloes
and the bias as determined by the peak-background split model
\begin{eqnarray}
  n&\to&n-\left(1+\frac{2f_{NL}}{\alpha(k)}\right)
    \frac{dn}{d\delta_c}\delta_l\,, \\
  b&=&1-\left(1+\frac{2f_{NL}}{\alpha(k)}\right)\frac{d\ln n}{d\delta_c}\,.
\end{eqnarray}
Because of the $k^2$ term in $\alpha(k)$, the bias diverges to
large-scales, giving a detectable signature with amplitude dependent
on $f_{NL}$.

Unfortunately, the large-scale clustering signal is one of the most
difficult to measure, not just because large volumes are required to
reduce sample variance. In fact, observational systematics can
strongly affect the clustering on large-scales. For the SDSS,
variations in stellar density and seeing for imaging data can lead to
target density fluctuations giving rise to large-scale power in
spectroscopic galaxy samples that is degenerate with the $f_{NL}$
signature \shortcite{ross11,ross12}. Current galaxy survey
measurements of local $f_{NL}$ are consistent with $f_{NL}=0$, with
contraints from galaxy surveys giving measurements of $f_{NL}\pm21$ at
$1\sigma$ \shortcite{giannantonio13}, which should be compared with
those from the CMB, which are of order $\pm5$ \shortcite{planck_ng}.

\subsection{Summary}

\begin{figure}
  \begin{center}
  \includegraphics[width=0.7\textwidth]{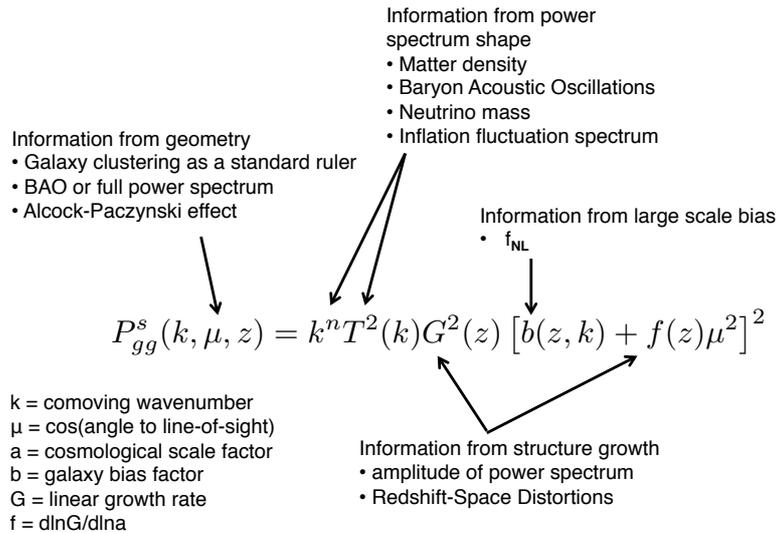}
  \end{center}
  \caption{Summary of the physics encoded in the observed large-scale
    galaxy clustering signal, as described by the linear power spectrum.}
  \label{fig:Lin_pk_physics}
\end{figure}

A summary of the physics encoded within the linear galaxy power
spectrum is presented in Fig.~\ref{fig:Lin_pk_physics}. This provides
a simple model for the power spectrum, showing the measurements that
can be made from each component.

\section{Making cosmological-model inferences}  \label{sec:bayes}

Modern observational cosmology relies on combining multiple
observations to make inferences about the correct cosmological
model. It uses Bayes theorem to do this. Suppose we have a hypothesis
to be tested $H$, a set of things assumed to be true $I$ (for example
the model of the observations), and a set of data $d$, then Bayes
theorem gives that
\begin{equation}  \label{eq:bayes}
  p(H|d,I)=\frac{p(d|H,I)p(H|I)}{p(d|I)}\,,
\end{equation}
where $L(H)\equiv p(d|H,I)$ is the sampling distribution of data,
often called the Likelihood, $p(H|I)$ is the prior, $p(d|I)$ the
normalisation, and $p(H|d,I)$ the posterior probability. 

The hypothesis to be tested is often a vector of parameters $\theta$,
which can be split into interesting $\phi$ (e.g. the dark energy
equation of state $w(z)$), and uninteresting $\psi$ (e.g. the
parameters of the smooth fit to the broad-band power when extracting
the BAO signal) parameters. Parameter inference is performed as
\begin{equation}  \label{eq:param_inf}
   p(\theta|d,I)\propto\int L(\phi,\psi)p(\phi,\psi|I)d\psi\,,
\end{equation}
in order to marginalise over the nuisance parameters, while retaining
information about the interesting parameters. Often we wish to know
the probability distribution for each parameter in turn having
marginalised over all others.

The primary concern when making parameter inferences is keeping a
handle on the physics that is being used, and allowing for all of the
potential sources of systematic error. When multiple data sets are
combined it becomes difficult to follow systematics through the
procedure, or to understand which parts of the constraints are coming
from which data set. In addition, a prior is required for the
hypothesis, whose impact can sometime be large, such that constraints
come from this, rather than the data. 

\subsection{Exploring parameter space}

Often, the list of parameters is long, and thus multi-parameter
likelihood calculations would be computationally expensive using
grid-based techniques. Consequently, fast methods to explore parameter
spaces are popular, particularly the Markov-Chain Monte-Carlo (MCMC)
technique, which is commonly used for such analyses. While there is
publicly available code to calculate cosmological model constraints
\shortcite{lewis02}, the basic method is extremely simple and relatively
straightforward to code.

The MCMC method provides a mechanism to generate a random sequence of
parameter values whose distribution matches the posterior probability
distribution. These sequences of parameter, or ``chains'' are commonly
generated by an algorithm called the Metropolis algorithm
\shortcite{metropolis53}: given a chain at position ${\bf x}$, a
candidate point ${\bf x_p}$ is chosen at random from a proposal
distribution $f({\bf x_p}|{\bf x})$ - usually by means of a random
number generator tuned to this proposal distribution. This point is
always accepted, and the chain moves to point ${\bf x_p}$, if the new
position has a higher likelihood. If the new position ${\bf x_p}$ is
less likely than ${\bf x}$, then we must draw another random variable,
this time with uniform density between 0 and 1. ${\bf x_p}$ is
accepted, and the chain moves to point ${\bf x_p}$if the random
variable is less than the ratio of the likelihood of ${\bf x_p}$ and
the likelihood of ${\bf x}$. Otherwise the chain ``stays'' at ${\bf
  x}$, giving this point extra weight within the sequence. In the
limit of an infinite number of steps, the chains will reach a
converged distribution where the distribution of chain links are
representative of the likelihood hyper-surface, given any symmetric
proposal distribution $f({\bf x_p}|{\bf x})=f({\bf x}|{\bf x_p})$
(see, for example, \shortciteNP{roberts96}).

It is common to implement dynamic optimisation of the sampling of the
likelihood surface (see other articles in the same volume as
\shortciteNP{roberts96} for examples), performed in a period of
burn-in at the start of the process. One thorny issue is convergence
-- how do we know when we have sufficiently long chains that we have
adequately sampled the posterior probability. A number of tests are
available \shortcite{gelman92,verde03}, although it's common to also
test that the same result is obtained from different chains started at
widely separated locations in parameter space.

\subsection{Model selection}

A lot of the big questions to be faced by future experiments can be
reduced to understanding whether a simple model is correct, or whether
additional parameters (and/or physics) are needed. For example, a lot
of future experiments are being set up to test whether the simple
$\Lambda$CDM model is correct. A model with more parameters will
always fit the data better than a model with less parameters (provided
it replicates the original model), while removing parameters may or
may not have a large impact on the fit. The interesting question is
whether the improvement allowed by having an extra free parameter is
sufficient to show that a parameter is needed?  Bayes theorem,
Eq.~(\ref{eq:bayes}), can test the balance between quality of fit and
the predictive ability, potentially providing an answer to these
problems.

The Bayesian evidence is defined as the average of the Likelihood
under the prior, given by $p(d|I)$ in Eq.~(\ref{eq:bayes}). Splitting
into model and parameters,
\begin{equation}
  p(d|I)=\int_\theta p(d|\theta,M)p(\theta|M)d\theta\,.
\end{equation}
The Bayes factor is defined as the ratio of the evidence for two
models
\begin{equation}
  B_{01}=\frac{p(d|I_0)}{p(d|I_1)}\,.
\end{equation}
The strength of the evidence that the Bayes factor gives is often
quantified by the Jeffries scale, which requires $B_{01}>2.5$ to give
moderate evidence for an additional parameter, and $B_{01}>5.0$ for
strong evidence. 

One problem with this approach is that the results depends on the
prior that is placed on the new parameter. For a wide prior covering
regions in parameter space that are unlikely, the average likelihood
and Bayes factor are reduced. A tight prior around the best-fit
position will lead to an increase in the Bayes factor. Because of
this, many variations on this test for new parameters have been
proposed (e.g. review by \shortciteNP{trotta08}).

\section{Future surveys}

A number of galaxy and Ly-$\alpha$ surveys are planned for the future,
in order to build up a picture of the distance-redshift relation over
a wide range of redshifts.

\subsection{The next 5 years}  \label{sec:future1}

The Sloan telescope will start to perform a spectroscopic galaxy
survey in 2014, called the extended Baryon Oscillation Spectroscopic
Survey ({\bf eBOSS}: http://www.sdss3.org/future/). This is the
cosmological survey within SDSS-IV, a six-year program using existing
hardware and an updated redshift measurement software pipeline. eBOSS
will provide distance measurements with BAO in the redshift range
$0.6<z<2.5$, approximately equivalent to a 1\% distance measurement at
$z=0.8$, two 2\% measurements at higher redshifts $z\sim0.9$ and
$z\sim1.5$ and a 1.5\% measurement at $z=2.5$ from clustering within
the Ly-$\alpha$ forest observed in the spectra from distant
quasars. These measurements will be complemented with RSD
measurements.

The Dark Energy Survey ({\bf DES}: www.darkenergysurvey.com) is using
a new 520-megapixel camera on the NOAO CTIO 4-meter Blanco telescope
to perform a 5000\,deg$^2$ multi-colour (grizY) imaging survey of the
southern hemisphere. When combined with VISTA data, this will provide
photometric redshifts for ~180,000,000 galaxies out to $z=1.5$. The
project is designed so that this catalogue can be used to measure the
equation of state of dark energy through the combination of
measurements from cluster number densities, weak lensing, supernovae
and galaxy clustering. Survey operations have recently started (Autumn
2013), with a 5-year baseline duration. Because only photometric
redshifts are available, only angular BAO will be measurable, and when
the data is combined we expect DES to provide a $\sim$2\% measurement
of the BAO distance scale at $z=1$.

Over 3 years, the Hobby-Eberly Telescope Dark Energy Experiment ({\bf
  HETDEX}) will take spectra and measure redshifts for 800,000
Ly-$\alpha$ emitting galaxies at $1.9<z<3.5$ over
420\,deg$^2$. Predictions for this survey are provided in
\shortciteN{greig13}, which suggest that a 1.2\% distance measurement
at $z=2.7$ is possible.

Two collaborations, Physics of the Accelerating Universe ({\bf PAU})
and Javalambre Physics of the Accelerating Universe Astrophysical
Survey ({\bf J-PAS}), are building new imaging cameras with the aim of
performing narrow-band imaging surveys, where excellent photometric
redshift can be obtained for of order 2-10,000,000 galaxies. PAU will
be based on the William Herschel Telescope (WHT), whereas J-PAS will
use a new dedicated 2.5m telescope in the South of Spain. Both surveys
envisage starting in 2014, and running for $\sim$5 years. Even using
narrow-band filters, the degradation due to having photometric
redshifts means that these surveys will be superseded by the next
generation of spectroscopic redshift surveys.

\subsection{5-20 years time} \label{sec:future2}

The next step for ground-based surveys is to place and utilise a new
multi-object spectrograph, capable of obtaining many thousands of
spectra simultaneously, on a 4m-class telescope. One project, called
Dark-Energy Spectroscopic Instrument ({\bf DESI}), will fit the Mayall
telescope with a new $\sim$5000-fibre spectrograph covering a
2--3\,deg diameter field. DESI is the most ambitious extra-galactic
ground-based survey proposed to date, given the fibre number, field of
view and fibre reconfiguration time \shortcite{schlegel11}. Survey
operations would be expected to commence in 2018. Other multi-object
spectrographs being designed for 4m telescopes include WEAVE
\shortcite{dalton12} and 4MOST \shortcite{dejong12}.

The Subaru telescope Prime Focus Spectrograph ({\bf PFS}) is a new
spectrograph due to have first light in 2017. It will have 2400
fibres, covering a 1.3\,deg diameter field of view
\shortcite{takada12}. The current cosmology survey design envisages a
survey over 1500 deg$^2$, yielding 3,000,000 galaxies over
$0.6<z<2.4$.

Looking beyond these ground-based surveys, the ESA {\bf Euclid}
mission, with a nominal launch date of 2019, will revolutionise
observational cosmology by conducting imaging and spectroscopic
surveys that are approximately two orders of magnitude better than
those currently available \shortcite{laureijs11}.

\subsection{Fisher methods}

It is common to make predictions for future surveys, in order to
compare them to each other and optimise them individually, using
Fisher methods. From Section~\ref{sec:bayes}, we have the Likelihood
$L(H)$. Then the Fisher matrix is defined as
\begin{equation}
  F_{ij}\equiv\left\langle\frac{\partial^2\mathcal{L}}{\partial\theta_i\partial\theta_j}\right\rangle\,,
\end{equation}
where $\mathcal{L}\equiv -\ln L$. For a multi-variate Gaussian
distribution,
\begin{equation}
  F_{ij}=\frac{1}{2}\left[
    C^{-1}\frac{\partial C}{\partial\theta_i}
    C^{-1}\frac{\partial C}{\partial\theta_j}\right]
    +\frac{\partial\mu^T}{\partial\theta_i}C^{-1}
    \frac{\partial\mu^T}{\partial\theta_j}\,,
\end{equation}
where $C$ is the covariance matrix, and $\mu$ the model for the
parameters being tested (e.g. power spectrum band-powers).  The
Cramer-Rao inequality shows that the diagonal elements of the Fisher
matrix give the best model error we can hope to achieve
\begin{equation}
  \Delta\theta_i\ge(F_{ii})^{-1/2}\,.
\end{equation}                                   
For a sample with constant number density within a volume $V$, the
error on the average power spectrum within a k-volume $V_k$
(e.g. \shortciteNP{FKP}) is
\begin{equation}
  \langle P({\bf k}) P({\bf k}')\rangle
    =\frac{1}{VV_k}\left[P(k)\delta_D({\bf k}-{\bf k'})+1/\bar{n}\right]^2\,.
\end{equation}
For a sample with varying density over the region surveyed, we can
define an effective volume
\begin{equation}
  V_{eff}\equiv\int\left[\frac{\bar{n}({\bf r})P(k)}{1+\bar{n}({\bf
          r})P(k)}\right]^2\,dr^3\,.
\end{equation}
If the survey is limited by the total number of redshifts that can be
obtained, then this has a maximum where $\bar{n}P=1$. For future
surveys we will be limited by the sky area we can observe, the
field-of-view of the telescope, or the redshift range we can test with
a given spectrograph, so we often have $\bar{n}P\ne1$. Based on the
effective volume,
\begin{equation}
  \langle P({\bf k})^2\rangle =2\frac{P(k)^2}{V_{eff}V_k}\,.
\end{equation}
This gives a Fisher matrix that can be written, if we now integrate
over many small shells $V_k$ \shortcite{tegmark97},
\begin{equation}  \label{eq:F_pk}
  F_{ij}=\frac{1}{2}\int\frac{d^3k}{(2\pi)^3}
    \left(\frac{\partial\ln P}{\partial\theta_i}\right)
    \left(\frac{\partial\ln P}{\partial\theta_j}\right)
    V_{eff}(k)\,.
\end{equation}
This is the starting point for designing and optimising future
surveys. Code that uses these ideas and isolates the BAO signal is
provided by \shortciteN{seo07}, while code for RSD measurements is
given by \shortciteN{white09}. Using these codes, we can predict
measurement errors on the measured distance scale, or on
$f\sigma_8$. Starting either from these measurements or
Eq.~\ref{eq:F_pk}, we can create Fisher matrices spanning the
parameters of complete cosmological models. 

In order to provide a simple statistic with which to compare
experiments, The Dark Energy Task Force (DETF;
\shortciteNP{albrecht06}) parametrized the equation of state $w(z)$ of
dark energy using two parameters $w_0$ and $w_a$:
\begin{equation}
  w(z)=w_0+\frac{z}{1+z}w_a\,.
\end{equation}
They then defined their Figure-of-Merit (FoM) to be proportional to
the inverse area of the error ellipse in the $w_0$--$w_a$ plane
\begin{equation}
  FoM_{\rm DETF}=\left[{\rm det}\,C(w_o,w_a)\right]^{-1/2}\,,
\end{equation}
where $C(w_o,w_a)$ is the covariance matrix after marginalising over
all of the other cosmological parameters. A larger $FoM_{\rm DETF}$ is
desired as it corresponds to a smaller error ellipse. Note that
$FoM_{\rm DETF}$ only cares about two parameters: an experiment that
measures $w(z)$ very accurately but only over a narrow redshift range
will score well. As the true dark energy equation-of-state is unknown,
so $FoM_{\rm DETF}$ is not necessarily the best statistic with which to
compare the discovery potential of multiple projects. Other options
that attempt to alleviate these problems are available
(e.g. \shortciteNP{albrecht09}).

\subsection{Predictions for future surveys}

\begin{figure}
  \begin{center}
  \includegraphics[width=0.7\textwidth]{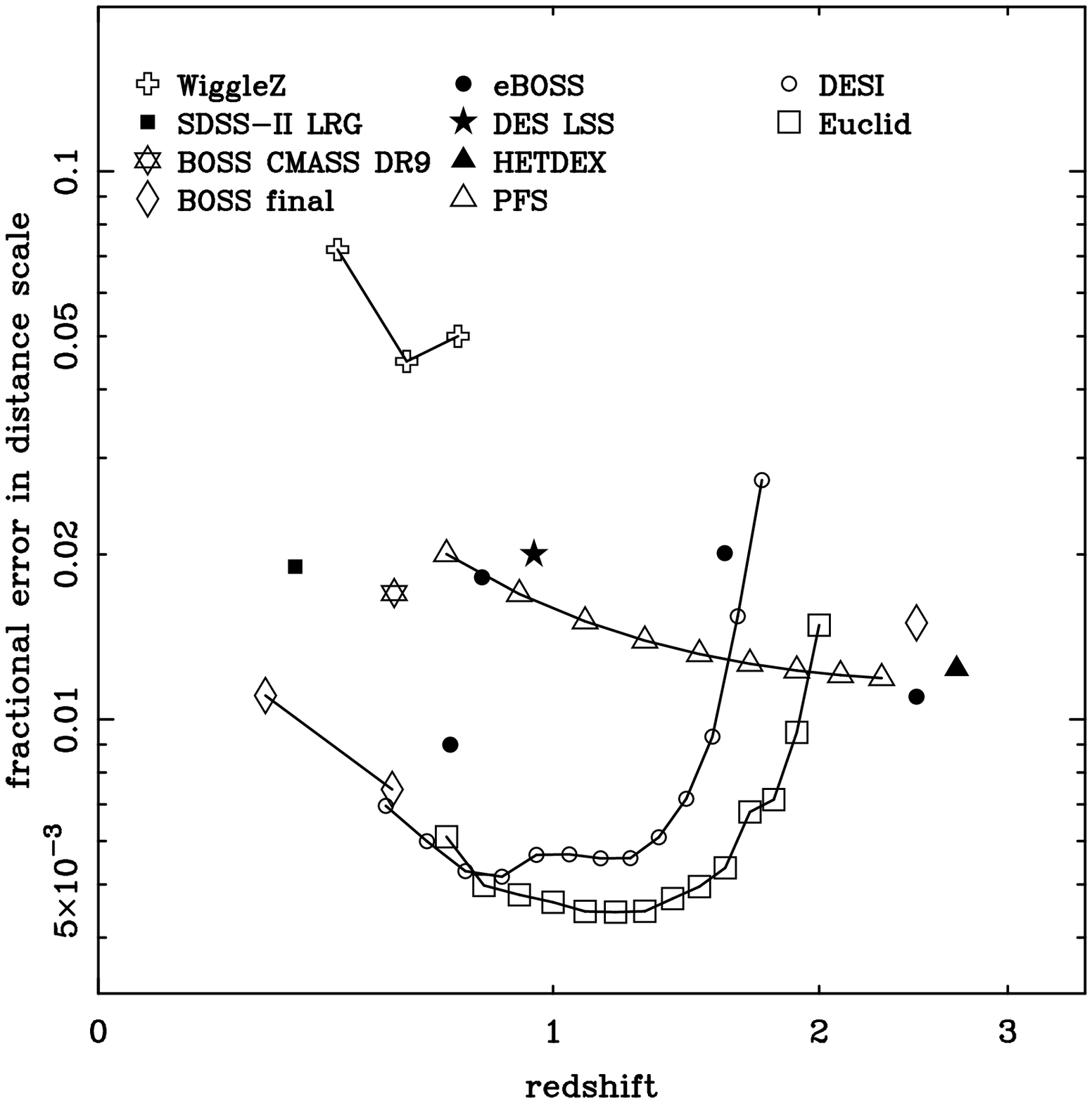}
  \end{center}
  \caption{Predictions for BAO measurements from the surveys
    introduced in Sections~\ref{sec:future1} \&~\ref{sec:future2},
    calculated using the code of \protect\shortciteN{seo07}. }
  \label{fig:cmpr_surveys}
\end{figure}

Predicted BAO measurements for the surveys introduced in
Sections~\ref{sec:future1} \&~\ref{sec:future2}, are presented in
Fig.~\ref{fig:cmpr_surveys}. The final BOSS galaxy survey is predicted
to provide multiple sub-percent BAO measurements out to redshifts of
$z\sim0.6$. Surveys that are starting now will provide 1--2\%
measurements over a wider redshift range, including percent level
measurements from the Ly-$\alpha$ forest. The next generation of
surveys including DESI and Euclid will provide multiple sub-percent
level measurements within redshift bins of width $\Delta z=0.1$ over
the full redshift range $0.6<z<2.0$.

In order to realise the power of future surveys, we will need
systematic errors that are significantly below the level of the
statistical errors shown in Fig.~\ref{fig:cmpr_surveys}. To ensure
this, all aspects of the analysis will need to be evaluated for
potential errors, including the likelihood calculation, and model to
be fitted to the data. Provided this is achieved then the future
surveys will provide a stunning new insight into the evolution of the
Universe and the forces driving it.

\vspace{.5cm}
\leftline{\bf Acknowledgements}

I would like to thank to organisers of the schools for inviting me to
give this series of lecture and to write these notes. I acknowledge
support from the UK Science \& Technology Facilities Council (STFC)
through the consolidated grant ST/K0090X/1, and from the European
Research Council through the ``Starting Independent Research'' grant
202686, MDEPUGS. Although I retain the sole right to any mistakes, I
would like to thank Angela Burden and Cullan Howlett for proof-reading
this draft.

\bibliographystyle{OUPnamed}
\bibliography{Percival_lecture_notes.bib}

\end{document}